\documentclass[12pt]{article}
\usepackage[utf8]{inputenc}
\usepackage{lmodern}
\usepackage[T1]{fontenc}

\usepackage{graphicx}
\usepackage{amsmath,amsthm,amssymb}
\usepackage{bbm}
\usepackage{enumitem}
\usepackage{url}
\usepackage{setspace}
\usepackage{nicefrac}
\usepackage{dsfont} 

\usepackage[margin=3.2cm]{geometry}
\usepackage{titlesec}
\titleformat*{\section}{\large\bfseries}
\titleformat*{\subsection}{\bfseries}

\def\ds{\displaystyle}

\newtheorem{thm}{Theorem}
\newtheorem{prop}[thm]{Proposition}

\newtheorem{cor}[thm]{Corollary}
\newtheorem{example}{Example}

\begin{document}

\thispagestyle{empty}
\def\thefootnote{\fnsymbol{footnote}}

\begin{center}
{\large \bf Market allocations under conflation of goods}\footnote{We thank Achille Basile, Maria Gabriella Graziano, Salvatore Modica, Marialaura Pesce, and seminar audiences for their comments. We are grateful to the referees for their insights and to Rakesh Vohra for his first-class editorial stewardship. Financial support under PRIN 20222Z3CR7 and GRINS PE00000018 is acknowledged. An Online appendix is available at \url{https://mlicalzi.github.io/research/MACG-Online-Appendix.pdff}.}
\end{center}

\setcounter{footnote}{0}
\def\thefootnote{\alph{footnote}}

\vspace*{.1cm}
\begin{center}
\begin{tabular}{ccc}
{\sc Niccol\`{o} Urbinati} & $\quad$ & {\sc Marco LiCalzi} \\
{\small Universit\`{a} Ca' Foscari Venezia} & &  {\small Universit\`{a} Ca' Foscari Venezia} \\
{\small\tt niccolo.urbinati@unive.it} & & {\small\tt licalzi@unive.it}\\
\end{tabular}
\end{center}
\vspace*{0.2cm}
\begin{center}
April 2025
\end{center}
\vspace*{0.4cm}

{\small
\noindent {\bf Abstract.} We study competitive equilibria in exchange economies when a continuum of goods is conflated into a finite set of commodities. The design of conflation choices affects the allocation of scarce resources among agents, by constraining trading opportunities and shifting competitive pressures. We demonstrate the consequences on relative prices, trading positions, and welfare.


}

\setcounter{footnote}{0}
\def\thefootnote{\arabic{footnote}}

\bigskip

\section{Introduction}

The intuitive notion of a commodity is ambiguous: ``Should two apples of different sizes be considered two units of the same commodity?'' (Geanakoplos, 1989, p.~44). Economic theory relies on the idealized notion of an Arrow-Debreu commodity, whose objective description includes all relevant characteristics, including geographic and temporal location.

In principle, the Arrow-Debreu commodities of an economy are so finely identified that no further refinements can yield Pareto-improving allocations. In practice, markets for Arrow-Debreu commodities are usually so thin that they are rarely traded. Real 
``commodity markets require design of both the marketplace procedures and the commodities themselves'' (Roth 2018, p.~1611).  After hinting at the uncountable expanse of theoretical goods, even Debreu (1959, p.~32) abruptly informs the reader that ``it is assumed that there is only a \emph{finite} number of distinguishable commodities'' (emphasis added). His assumption mutes that tradable commodities are acknowledged by convention or created by design. 

There are different ways to establish a commodity market. We focus on \emph{conflation} by which ``similar but distinct products are treated as identical in order to make markets thick or reduce cherry-picking'' (Levin and Milgrom, 2010, p.~603). Here is an example. The people of a village cultivate a vineyard on the banks of a long river. The quality of the grapes depends on the position of the vines: one obtains different types of wine by pressing grapes grown on different vines. By custom, the villagers blend their grapes and trade at most two types of wine. Their technology separates the grapes growing on the lowlands ($L$) from those growing on the hills ($H$). The position $\theta$ of the boundary between $L$ and $H$ defines the conflation of grapes into two wines.

Conflation is usually achieved by setting standards, similarly to choosing the position $\theta$ in our example. Conflation (and deconflation) are techniques for product definition involved in the creation or design of a commodity market (Milgrom, 2011). The frequent batch auctions (FBA) proposed in Budish (2015) can be interpreted as conflating trading opportunities over a time interval to a single instant.\footnote{\ Chauvin~(2024) studies trade-restricted competitive equilibria.} Conflation takes place at the market level when several goods are made available only as a single package (Adams and Yellen, 1976).

This paper considers exchange economies where a continuum of goods is conflated into a finite number of tradable commodities. We assume that the goods are partitioned by means of a \emph{classification} that associates any good in the endowment to exactly one commodity.

A classification arranges the existing endowment into commodities and defines what can be traded. This differs from changing the set of available goods or introducing a new commodity into an economy. Gilles and Diamantaras~(2003) studies a general equilibrium model including the costly institutional arrangements required by the arrival of new tradable commodities. Stokey~(1988) analyzes a dynamic general equilibrium model that sustains growth by introducing new and better products. Moreover, we assume classifications as given and do not analyze how they emerge or are selected. Sprumont~(2004) takes a first step towards endogenizing classifications, using an axiomatic approach where agents' preferences reveal subsets of characteristics that are sufficiently homogeneous and distinct to deserve being singled out as commodities.

We imagine a price-based market economy as an institution that binds agents' final allocations to the extant classification. Viewed as an intangible good, the classification is non-excludable and non-rivalrous: it may be interpreted as an abstract public good in the sense of Mas-Colell (1980) and Diamantaras and Gilles (1996).

Our main point is that the choice (or design) of a classification directly affects the outcome. Configuring the set of tradable commodities shifts the strength of agents' demands over the underlying goods, inducing significant changes in how trades equilibrate. Competitive (equilibrium) allocations depend on the classification supporting them.

This paper studies and compares competitive allocations generated by different classifications over the same endowment of goods. The set of price-based competitive allocations in our model depends continuously on the choice of the classification, and therefore, agents receive similar payoffs in economies using similar classifications. We show that, in equilibrium, the relative prices of two commodities depend on the classification of other (distinct) commodities; moreover, the classification of other commodities may affect the allocation of the two commodities even if their relative prices stay the same. 

Importantly, any classification defines a different economy: a competitive allocation is Pareto-optimal given its underlying classification, but changing the classification may create another economy where every agent is better off. As an abstract public good, a classification tacitly induces externalities that the competitive equilibria cannot price or otherwise take into account. We argue that revising a classification induces new competitive allocations (with different externalities) that may increase efficiency.

The literature on incomplete (financial) markets has shown that introducing a new security may lead to competitive equilibria that are Pareto-inferior; see Hart (1975, Section~6) or Cass and Citanna (1998). Our model, instead, demonstrates for (real) exchange economies that adopting a finer classification can lead to competitive equilibria that are never Pareto-improving, and may even reduce social welfare. The common theme is that increasing trading opportunities is not necessarily advantageous, even in the absence of transaction costs.

\section{The model}\label{sec2}

Consider a society endowed with a continuum of goods to be allocated over a finite set of $n$ agents. The agents have preferences on consumption and hold compatible claims on society's endowment. It is infeasible (or impractical) to trade over a continuum, but society can classify goods and conflate them into a finite number of commodities. The adoption of a classification defines which commodities are tradable in the economy. The classification is shared by all agents and binds their choices: an agent can demand and be allocated only bundles of tradable commodities.

\paragraph{Goods and commodities.} 

Let $\mathcal{I}$ be the space of goods. Each element $t$ in $\mathcal{I}$ corresponds to a complete description of the attributes of the good. Given an algebra of sets $\mathcal F$ on $\mathcal{I}$, we interpret every subset $C$ in $\mathcal F$ as the \emph{commodity} generated by the conflation of the goods in $C$.\footnote{\ From a technical point of view, this is consistent with the approach that goods aggregate characteristics (Lancaster, 1966). It suffices to redefine primitives and assume that goods are measurable sets of characteristics; see Mas-Colell~(1975) and Jones~(1984).} (In the sequel, any subset $C$ belongs to $\mathcal F$ and refers to a commodity.) 

The endowment of goods available to the economy is described by a positive measure $\omega$ on $\mathcal{I}$, normalized to $\omega(\mathcal{I})=1$. Every positive function $b$ in $L^1_+(\omega)$ describes a \emph{bundle} of goods. If the goods in $C$ are conflated into a commodity, the total amount available is the quantity $\omega(C)$ while $\int_C b\,d\omega$ is the amount provided by the bundle $b$.

\paragraph{Agents.}

There are $n$ agents. The preferences of agent $i$ over bundles of goods are represented by a utility function $U_i$ on $L^1_+(\omega)$ that is concave, monotone,\footnote{\  That is, $U_i(f) \geq U_i(g)$ for any $f,g\in L^1_+(\omega)$ with $f(t )\geq g(t)$ almost everywhere.} and norm-continuous. For convenience, every agent $i$ has a proportional\footnote{\ This proportionality assumption implies that our exchange economies may be interpreted as Fisher market models; see Vazirani~(2007). Alternatively, one may assume that every agent is allocated a budget.} claim $\kappa_i>0$ on the economy's endowment, and we assume $\sum_{i=1}^n\kappa_i=1$.

\paragraph{Classification of goods.} 

A \emph{classification} $\pi$ is a partition of $\mathcal{I}$ into a finite number of subsets. Each element $C$ in $\pi$ is interpreted as a conflation of the goods in $C$ into a tradable commodity. Conflation is technologically irreversible, as is the case when mixing wheat in adherence to a market standard or blending grapes into a wine with a protected designation of origin.\footnote{\ This mutes concerns about cherry-picking or asymmetric information, but other interpretations are possible. A recurrent suggestion is to imagine that, after trading has closed, a lottery over the commodity unpacks its underlying goods, with agents computing expected utilities from the tradable lotteries.} The partition $\pi$ defines which commodities are tradable in the economy.

A \emph{unit} of a commodity $C$ consists of a bundle $\frac{1}{\omega(C)}\mathbbm{1}_C$ of goods chosen uniformly from $C$. Given a (finite) classification $\pi$, a \emph{tradable bundle} specifies quantities for each of the commodities in $\pi$ and is described by a vector $x=(x_C)$ in $\mathbb{R}^\pi_+$, where $x_C$ is the quantity of goods in $C$. Correspondingly, a tradable bundle $x$ is represented by the simple function $\sum_{C\in\pi}\frac{x_C}{\omega(C)}\mathbbm{1}_C$.

\paragraph{Exchange economy after a classification.}

Consumers can demand or exchange only tradable bundles. Thus, an instantiation of the model combines the identification of tradable commodities via $\pi$ and a compatible allocation of the goods. We represent it as a configuration $\langle \pi,(x^i) \rangle$, where $\pi$ is the shared classification and each $x^i$ in $\mathbb{R}^\pi_+$ is the tradable bundle assigned to agent $i$. A configuration $\langle \pi,(x^i) \rangle$ is \emph{feasible} if the total amount of allocated commodities does not exceed their initial availability: i.e., if $\sum_{i}x_C^i\leq\omega(C)$ for every $C$ in $\pi$.

The classification $\pi$ defines an exchange economy $\mathcal{E}(\pi)$ over a (constrained) set of tradable bundles in $\mathbb{R}^\pi_+$. Every agent is endowed with the vector $e^i=(\kappa_i\omega(C))$ in $\mathbb{R}^\pi_+$ and evaluates the tradable bundles by the function
$$V_i(\pi,x)=U_i\left(\sum_{C\in\pi}\frac{x_C}{\omega(C)}\mathbbm{1}_C\right),$$
that is the restriction of $U_i$ to the set of simple functions measurable with respect to $\pi$. The function $V_i(\pi,\cdot)$ inherits continuity, concavity, and monotonicity from the function $U_i$. Because agents' preferences do not depend on the classification, there are no spurious effects driving our results.

A competitive equilibrium in the economy $\mathcal{E}(\pi)$ is a pair $\langle p,(x^i)\rangle$ formed by a price vector $p\in\mathbb{R}^\pi_+$ and a feasible allocation $(x^i)$ such that: $(i)$ $p\cdot x^i\leq \kappa_i {\sum_C} \,p_C\, \omega(C)$ for all $i$, and $(ii)$ if $V_i(\pi,y)>V_i(\pi,x^i)$ for some $\pi$-bundle $y$ then $p\cdot y>\kappa_i\sum_{C}\,p_C\,\omega(C)$. We say that a configuration $\langle \pi,(x^i)\rangle$ is \textit{competitive} if $(x^i)$ corresponds to a competitive equilibrium in $\mathcal{E}(\pi)$.

In short, imposing a classification $\pi$ reduces the economic interaction over a continuum of goods to a finite-dimensional economy $\mathcal{E}(\pi)$ where agents have concave, monotone, and continuous preferences and strictly positive endowments. It follows that for any classification $\pi$ a competitive equilibrium in $\mathcal{E}(\pi)$ exists.

In the following, we make two assumptions for the economy of exposition. First, $\mathcal{I}$ is the unit interval and $\mathcal F$ is the Borel $\sigma$-algebra. Second, the classification $\pi$ consists of intervals of positive measure.

\subsection{Notable examples}

The model postulates primitive preferences for bundles of goods in the infinite-dimensional space $L^1_+(\omega)$, and uses those to derive preferences over the finite-dimen\-sional subspaces induced by a classification. The primitive preferences over goods are hard to pin down in closed form, while their restrictions over commodities (after classification) appear fairly simple. We exhibit some examples of preferences on $L^1_+(\omega)$ that, once adapted to a classification $\pi$, induce popular utility functions on the commodity space $\mathbb{R}^\pi_+$.

When the primitive utility function $U_ i: L_+^1(\omega)\to \mathbb{R}$ is linear, we can associate it with a measure $\nu_i$ on $\mathcal{I}$ such that $U_i(b)=\int b(t)\,d\nu_i(t)$ for every bundle $b$. By our assumptions on $U_i$, the measure $\nu_i$ turns out to be absolutely continuous with respect to $\omega$. We call $\nu_i$ the agent $i$'s \emph{evaluation measure} and interpret it as a description of the value that $i$ assigns to each conflation of goods. This agent ranks bundles in $\mathbb{R}^\pi_+$ by the linear function
$$V_i(\pi,x)=\sum_{C\in\pi}\left(\frac{x_C}{\omega(C)}\right)\nu_i(C).$$

Suppose instead that the primitive utility function is $U_i(b)=\exp\left(\int \log\left[b(t)\right]\,d\nu_i(t)\right)$, with $U_i(b)=0$ when $b(t)=0$ on a non-null set. Then the agent ranks bundles in $\mathbb{R}^\pi_+$ by the Cobb-Douglas utility function
$$V_i(\pi,x)=\prod_{C\in\pi} \left( \frac{x_C}{\omega(C)}\right)^{\nu_i(C)}.$$

Finally, if the primitive utility is $U_i(b)=\left(\int b(t)^\rho\,d\nu_i(t)\right)^{1/\rho}$ for some $0<\rho<1$, then the agent ranks bundles in $\mathbb{R}^\pi_+$ by the CES-type utility function
$$V_i(\pi,x)=\left[\sum_{C\in\pi} \left(\frac{x_C}{\omega(C)}\right)^\rho\right]^{1/\rho}.$$

Under these utility functions, agents have well-behaved preferences in any (finite-dimensional) economy $\mathcal{E}(\pi)$ induced by a classification $\pi$. In particular, when all agents have linear (Cobb-Douglas, CES) utility functions, equilibrium prices in every $\mathcal{E}(\pi)$ exist and are unique up to normalization, and thus agents receive the same utility at every competitive equilibrium in $\mathcal{E}(\pi)$. This observation generalizes as follows. 

Suppose that $U_i (b)=\int u_i\left[t,b(t)\right]\,dt$ for some integrable map $u_i\colon\mathcal{I}\times \mathbb{R} \to \mathbb{R}$ where $u(t,\cdot)$ is increasing, concave, and such that
$$\frac{\partial u(t,v)}{\partial v} \ge -v\frac{\partial^2u(t,v)}{\partial v^2}$$
with a strict inequality when the left-hand side is non-zero. (In particular, all the examples above have equivalent representations of this form.) By applying the Leibniz integral rule, one shows that the function $V_i(\pi,\cdot)$ satisfies the conditions of Example~17.F2 in Mas-Colell et al.~(1995). It follows that if all agents have preferences of this type, then in every economy $\mathcal{E}(\pi)$ their aggregate demands satisfy Gross Substitutability, and therefore equilibrium prices exist and are unique, up to normalization.

\section{Classification matters}\label{sec3}

Standard models of exchange take the notion of commodity as primitive. We depart from this and consider alternative classifications for the same underlying set of goods, and their effect on competitive equilibrium allocations. For readability, proofs are relegated to the appendix.
 
We first show that similar classifications lead to similar exchange economies, and hence to similar sets of competitive equilibria. If we imagine a classification as the output of setting standards based on cut-offs for grade or quality, then small changes in the cut-offs entail small changes in the set of competitive allocations. More generally, we emphasize that our model ensures that tiny changes in a classification cannot lead to dramatic changes in the equilibrium outcomes.

 To formalize this intuition, we endow the set $\Pi_{(\le k)}$ of the classifications using \emph{at most} $k$ intervals with a topological structure. Let $\sigma(\pi)$ denote the $\sigma$-algebra generated by a classification $\pi$ and set 
$$\delta_\omega(\pi,\pi') = \sup_{C \in \sigma(\pi)} \, \inf_{C^\prime \in \sigma(\pi^\prime)} \, \omega(C \triangle C^\prime).$$
Then the function $d_\omega(\pi,\pi') = \max \{\delta_\omega(\pi,\pi'),\delta_\omega(\pi',\pi)\}$ is a Hausdorff pseudo-metric for the set of classifications, induced by the measure of the symmetric difference between sets; see Boylan (1971).

\begin{prop}\label{prop-1}
$(\Pi_{(\le k)},d_\omega)$ is a compact space where two classifications have zero distance if and only if they coincide up to null sets. 
\end{prop}
 
In short, two classifications are essentially equivalent when their distance is zero. Recall from Section~\ref{sec2} that under our assumptions the set $\mathcal{W}(\pi)$ of competitive equilibria for an economy $\mathcal{E}(\pi)$ is not empty. We show that $\mathcal{W}(\pi)$ is compact and depends continuously on the choice of the classification $\pi$.

\begin{thm}\label{thm.0}
The competitive equilibria correspondence $\mathcal{W}$ is compact-valued and upper-hemicontinuous on the compact space $(\Pi_{(\le k)},d_\omega)$.
\end{thm}

We obtain as a corollary that if goods are allocated through competitive equilibria, then agents receive similar payoffs in economies based on similar classifications. Let $\Psi_i(\pi)$ be the set of utilities that an agent $i$ may obtain across all competitive equilibria for the economy $\mathcal{E}(\pi)$.

\begin{cor}
The competitive utilities correspondence $\Psi_i$ is compact-valued and upper-hemicontinuous on the compact space $(\Pi_{(\le k)},d_\omega)$.
\end{cor}

Continuity is crucial for comparative statics. If agents' utilities depend continuously on the classification, we can study not only \emph{if} the welfare changes with the classification, but \emph{how} it changes.

For instance, consider an economy based on a binary classification $\pi=\{A,B\}$. One may conjecture that increasing the number of commodities by one (e.g., by splitting $A$ into two intervals) alters the equilibrium utilities less than introducing a hundred new commodities. This is not the case. Because of continuity, if the new classification $\rho$ is sufficiently close to $\pi$ (in the topological sense), then the change in the agents' equilibrium utilities under $\pi$ and $\rho$ can be made arbitrarily small, regardless of the number of new commodities in $\rho$. See Section~\ref{opposed} for a simple exercise in comparative statics.

\subsection{Pareto-optimality and welfare theorems}

We turn to studying the efficiency of competitive configurations. We say that a configuration $\langle \pi,(x^i)\rangle$ \textit{Pareto-dominates} another configuration $\langle\rho,(y^i)\rangle$ if $V_i(\pi,x^i)\geq V_i(\rho,y^i)$ for all $i$, with a strict inequality for at least an agent $i$. Given a set of feasible configurations $\mathcal{F}$, a configuration $\langle \pi,(x^i)\rangle\in\mathcal{F}$ is \textit{Pareto-optimal} in $\mathcal{F}$ if there is no configuration in $\mathcal{F}$ that Pareto-dominates it. 

It is a simple observation that any competitive configuration is Pareto-optimal among those based on the same classification; that is, if $\langle \pi,(x^i)\rangle$ is competitive, then no feasible configuration of the type $\langle \pi,(y^i)\rangle$ Pareto-dominates it. This is a consequence of the first Welfare Theorem applied to the exchange economy $\mathcal{E}(\pi)$. 

A more interesting scenario opens when we compare allocations based on different classifications: then it is no longer true that any competitive configuration is Pareto-optimal. For instance, the competitive equilibrium under the trivial classification $\pi=\{\mathcal{I}\}$ is generically Pareto-dominated by many other finer competitive configurations. In the following example,  a competitive configuration is (strictly) Pareto-dominated by another competitive configuration with the same number of commodities, even if the relative prices for the commodities are the same.

\begin{example}\label{Pareto.1}
Consider a society where $\omega$ is the Lebesgue measure, and there are three agents who all have the same claim $\kappa =\nicefrac{1}{3}$. Agents' preferences are linear and respectively based on the evaluation measures
$$\nu_1(F)=3\omega\left(F\cap \left[0,\frac{1}{3}\right]\right) \>\mbox{ and }\> \nu_2(F)=\nu_3(F)= 3\omega\left(F\cap\left[\frac{1}{3},\frac{1}{2}\right)\right)+\omega\left(F\cap \left[\frac{1}{2},1\right]\right);$$
that is, agent $1$ cares only about goods in the first third of the interval (and is indifferent over them), while agents $2$ and $3$ value any good in $\left[\frac{1}{3},\frac{1}{2}\right)$ thrice as much as those in $\left[\frac{1}{2},1\right]$.

Consider the economy $\mathcal{E}(\pi)$ under the classification formed by the intervals $A=\left[0,\frac{1}{2}\right)$ and $B=\left[\frac{1}{2},1\right]$.  Agent $1$'s evaluations for the two tradable commodities are respectively $1$ and $0$, and thus he demands only the first commodity. Agents~$2$ and $3$ have identical evaluations for the commodities and demand whatever is cheaper. A competitive equilibrium has identical prices for the commodities: it assigns the $\pi$-bundle $x^1=\left(\frac{1}{3},0\right)$ to agent $1$, and the bundles $x^2=x^3=\left(\frac{1}{12},\frac{1}{4}\right)$ to agents $2$ and $3$. At the competitive configuration $\langle \pi,(x^i)\rangle$ the agents have utilities:
$$V_1(\pi,x^1)=\frac{2}{3},\ \ V_2(\pi,x^2)=\frac{1}{3},\ \ V_3(\pi,x^3)=\frac{1}{3}.$$

Consider the alternative classification $\rho=\{A^\prime,B^\prime\}$ where $A^\prime=\left[0,\frac{1}{3}\right)$ and $B^\prime=\left[\frac{1}{3},1\right]$. In the economy $\mathcal{E}(\rho)$, agent $1$ demands only commodity $A^\prime$ and agents $2$ and $3$ demand only commodity $B^\prime$. A competitive equilibrium has identical prices for the commodities: it assigns the $\rho$-bundle $y^1=\left(\frac{1}{3},0\right)$ to agent $1$, and the bundles $y^2=y^3=\left(0,\frac{1}{3}\right)$ to agents $2$ and $3$. In this case, agents have utilities:
$$V_1(\rho,y^1)=1,\ \ V_2(\rho,y^2)=\frac{1}{2},\ \ V_3(\rho,y^3)=\frac{1}{2}.$$
Clearly, the competitive configuration $\langle \rho,(y^i)\rangle$ Pareto-dominates the competitive configuration $\langle \pi,(x^i)\rangle$.
\end{example}

This example does not describe an isolated case. Whenever agents are not all identical, there is a ``rich'' set $O$ of classifications for which every competitive configuration based on a classification in $O$ is Pareto-dominated. In short, once classifications enter the picture, the first Welfare Theorem may fail in non-pathological circumstances.

\begin{prop}\label{prop:OpenSet}
Suppose that agents are not all identical and $k>1$. Then there is an open set $O$ of classifications in $\Pi_{(\le k)}$ whose competitive configurations are Pareto-dominated by some competitive configuration $\langle\hat{\pi},(\hat{x}^i)\rangle$ with $\hat{\pi}$ in $\Pi_{(\le k)}$.
\end{prop}

One can recast the notion of a rich set $O$ of Pareto-dominated competitive configurations in measure-theoretic terms; see Corollary~\ref{cor:NonZeroSet} in the online Appendix. 

The failure of Pareto optimality for a competitive configuration $\langle \pi,(x^i)\rangle$ follows because the classification $\pi$ constrains trading over goods only to commodities. As this restriction violates the assumption of universality of markets (Arrow, 1970), the first Welfare Theorem does not hold across different classifications. Classifications (or analogous restrictions on trade) may be crucial for the design of thick markets, but they come with a risk of efficiency loss.

Nonetheless, because agents' preferences are continuous, one can prove the existence of competitive configurations that are Pareto-optimal across different classifications.

\begin{thm}\label{thm.2}
The set of competitive configurations based on classifications in $\Pi_{(\le k)}$ has a Pareto-optimal configuration.
\end{thm}

This result states that there exists a Pareto-optimal configuration among all the classifications using at most $k$ commodities.  In general, if there is no upper bound on the number of commodities forming a classification, there may not exist a Pareto-optimal configuration, even within the set of competitive configurations. Example~\ref{Pathological} in the online Appendix exhibits a pathological case where every competitive configuration can be Pareto-improved by a classification based on a strictly larger number of commodities.

Perhaps surprisingly, while the first Welfare Theorem fails for competitive configurations based on different classifications, it is possible to prove a version of the second Welfare Theorem for competitive configurations. 

\begin{prop}\label{Prop.2}
If $\langle\pi,(x^i)\rangle$ is a Pareto-optimal configuration such that $x^i_C>0$ for every $i$ and every $C$ in $\pi$, then one can redefine agents' claims so that $\langle\pi,(x^i)\rangle$ is competitive.
\end{prop}

Proposition~\ref{Prop.2} ensures that an interior allocation that is Pareto-optimal can be cast as a competitive configuration after suitably modifying agents' claims. This aligns well with the classical version of the second Welfare Theorem, by which any interior Pareto-optimal allocation in an exchange economy is a competitive equilibrium for some suitable initial distribution of resources. 

\subsection{The relative scarcity of commodities}

In a competitive equilibrium, one often interprets the ratio of the prices of two commodities as an index of relative scarcity: given preferences and endowments, the greater is the ratio, the higher is the value attributed to the first commodity. We argue that this ratio is not an intrinsic property of the two commodities, because it depends on how other distinct commodities are classified.

The next example keeps two commodities fixed and studies how the ratio of their prices varies as we change the rest of the classification. Even if agents' evaluations of the two commodities remain constant, the ratio of their prices ranges over an interval that can be made arbitrarily large. This implies that knowing the ratio of the prices of two commodities is meaningless without a full description of the whole classification. 

\begin{figure}[ht]
\centering
\includegraphics[width=0.4\textwidth]{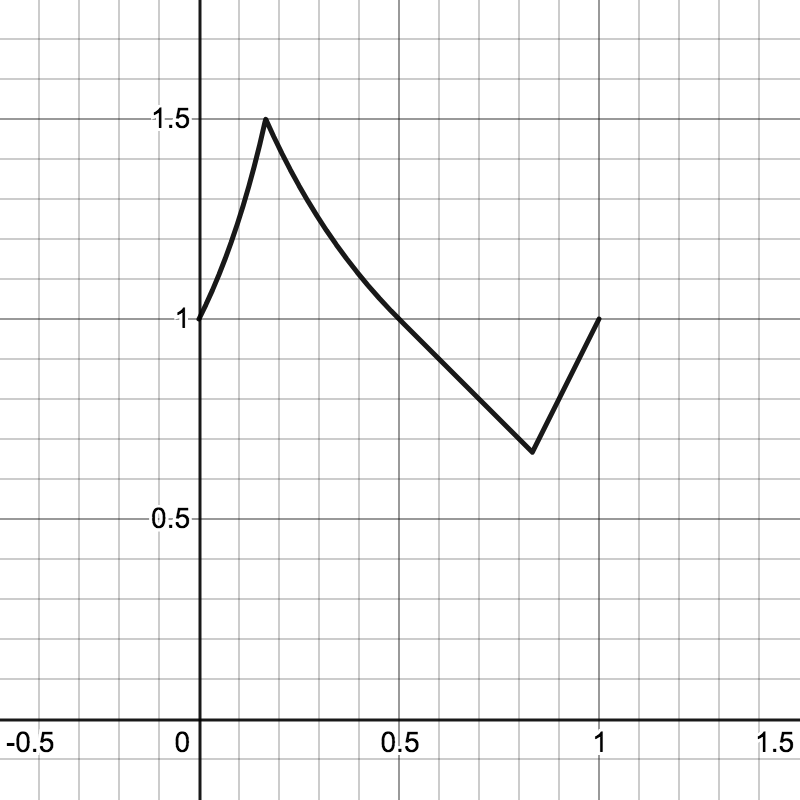}
\caption{Ratio of the equilibrium prices for commodities $A$ and $D$.}\label{Graph}
\end{figure}

\begin{example}\label{Ex.RS}
Consider a society where $\omega$ is the Lebesgue measure, and there are $2n$ agents who all have identical claims. There are two types of agents, forming groups of equal size. Agents' preferences are linear and based on the evaluation measures:
$$\nu_1(F)=2\omega\left(F\cap\left[0,\frac{1}{2}\right]\right) \quad\mbox{ and }\quad \nu_2(F)=2\omega\left(F\cap\left[\frac{1}{2},1\right]\right).$$
For every $t\in (0,1)$, let $\pi_t$ be the classification formed by the four intervals:
$$A=\left[0,\frac{1}{4}\right),\ \ B_t=\left[\frac{1}{4},\frac{1+2t}{4}\right),\ \ C_t=\left[\frac{1+2t}{4},\frac{3}{4}\right),\ \ D=\left[\frac{3}{4},1\right].$$
The two commodities $A$ and $D$ are acknowledged as tradable in any classification $\pi_t$, while the other two tradable commodities $B_t$ and $C_t$ depend on the choice of $t$. We claim that the ratio of equilibrium prices for the two commodities $A$ and $D$ depends on the threshold $t$.

For every $t$, let $p_t$ be a competitive price system in $\mathcal{E}(\pi_t)$ and let $\varphi(t)$ denote the ratio $p_t(A)/p_t(D)$. Note that the function $\varphi$ does not depend on how the $p_t$'s are chosen, because two competitive prices in $\mathcal{E}(\pi_t)$ must be proportional to each other. Assuming $p_t(D)=1$ for all $t$, we compute the equilibrium prices case by case.

Suppose $t\leq \frac{1}{2}$. Then commodities $A$ and $B_t$ are desirable only for agents in group $1$, $C_t$ is desirable for agents of both groups, and $D$ is desirable only for agents in group $2$. In equilibrium agents from group~$1$ demand commodities $A$, $B_t$ and $C_t$ as long as $t\leq \frac{1}{6}$, and demand only commodities $A$ and $B_t$ if $t>\frac{1}{6}$. On the other hand, agents from group~$2$ demand positive amounts of commodities $C_t$ and $D$. Because $p_t(D)=1$, the resulting equilibrium prices are:
$$p_t(A)=\begin{cases} 
	\frac{1}{1-2t}&\text{if }t \leq \frac{1}{6},\\
	\frac{2}{1+2t}&\text{if } t > \frac{1}{6},
	\end{cases}\qquad
p_t(B_t)=\begin{cases}
	\frac{1}{1-2t}&\text{if }t\leq \frac{1}{6},\\
	\frac{2}{1+2t}&\text{if } t > \frac{1}{6},
	\end{cases}\qquad
p_t(C_t)=\frac{1}{2(1-t)}.$$

Suppose instead $t\geq \frac{1}{2}$. The situation is symmetric to the above. In equilibrium, agents from group~$1$ demand commodities $A$ and $B_t$, while those from group~$2$ demand only commodities $C_t$ and $D$ if $t\leq \frac{5}{6}$, and may add commodity $B_t$ when $t\geq \frac{5}{6}$. The resulting equilibrium prices are:
$$p_t(A)=\begin{cases}
	\frac{3-2t}{2},&\text{if } t\leq \frac{5}{6},\\
	2t-1,&\text{if } t > \frac{5}{6},
	\end{cases}\qquad
p_t(B_t)=\begin{cases}
	\frac{3-2t}{4t}&\text{if }t\leq \frac{5}{6},\\
	\frac{2t-1}{2t}&\text{if } t > \frac{5}{6},
	\end{cases}\qquad
p_t(C_t)=1.$$
The function $\varphi(t)$ coincides with $p_t(A)$: it is increasing for $t\leq \frac{1}{6}$, decreasing for $\frac{1}{6} < t < \frac{5}{6}$, and increasing again for $t\geq \frac{5}{6}$. Its graph, plotted in Figure~\ref{Graph}, ranges from $\nicefrac{2}{3}$ to $\nicefrac{3}{2}$.
\end{example}

The index of relative scarcity for commodities $A$ and $D$ depends on the classification of tradable commodities. Moreover, for every threshold $t$ there exists $s$ such that $\varphi(t)=1/\varphi(s)$, so that $p_t(A)/p_t(D)=p_s(D)/p_s(A)$. Therefore, whatever holds about the equilibrium {price} of $A$ relative to $D$ in a classification is specularly true about $D$ relative to $A$ under a different classification.

Note that changes in the threshold $t$ affect the utilities that agents receive in equilibrium, and hence their relative welfare within the society (despite all agents having identical claims). In particular, the ratio $u^1(t)/u^2(t)$ of the equilibrium utilities that agents in the two groups $i=1,2$ receive in the economy $\mathcal{E}(\pi_t)$ has the same graph as the function $\varphi$ plotted in Figure~\ref{Graph}.

In Example~\ref{Ex.RS}, the ratio of the prices of two commodities $A$ and $D$ ranges over a closed interval as we change the classification and keep $A$ and $D$ fixed. One can modify the example and make the range of the interval arbitrarily large. However, because agents' preferences and the function $\varphi(t)$ are continuous and the set of classifications we consider is compact, the range must remain bounded.

\subsection{Economies with opposed preferences}\label{opposed}

This section demonstrates an example of comparative statics. (See Section~\ref{app-opp} of the online Appendix for an outline of the arguments and of the computations.) 

Assume that $\omega$ is the Lebesgue measure, and there are two agents with identical claims and linear preferences. The evaluation measure for agent $i=1,2$ is given by the integral of a positive density $f_i$ with full support over $\mathcal{I}$: 
$$\nu_i(F)=\int_F f_i(t)\,dt.$$
Agents have \emph{opposed preferences} when $f_1$ is decreasing and $f_2$ is increasing. Then there exists a single \emph{leveling point} $\theta$ in $\mathcal{I}$ such that $\nu_1\left( \left[0, \theta \right)\right) = \nu_2 \left( \left[ \theta,1\right]\right)$. 

For a motivating example, let us return to the vineyard in the introduction. Suppose that the grapes on higher ground mature faster and contain more sugar, leading to a higher alcohol content when used to produce wine. Assume two agents with opposed preferences about alcohol content in wine. The first agent likes better wine produced using grapes from the lowlands, while the second agent prefers wine based on grapes from the hills.

Arrange the intervals forming the classification $\pi = \{C_1,\dots,C_k\}$ in their natural (increasing) order. Let $C_{j^*}$ denote the commodity in $\pi$ that contains the leveling point $\theta$. Clearly, Agent~$1$ prefers all the commodities preceding $C_{j^*}$ more than Agent~$2$, who in turn likes the commodities following $C_{j^*}$ better than $1$. Thus, computing the final equilibrium allocation under $\pi$ reduces to studying how $C_{j^*}$ is allocated between 1 and 2.

Specifically, let 
$$C_\ell=\bigcup_{j<j^*} C_j  \quad\text{ and }\quad C_r = \bigcup_{j>j^*} C_j$$
denote the two commodities obtained by bundling the intervals on the left and on the right of  $C_{j^*}$, respectively. Because $\left[ \nu_1 (C_j) - \nu_2(C_j) \right] ( j^* - j) >0$ for any $j\ne j^*$, in equilibrium Agent~$1$ is allocated all the goods in $C_\ell$ and none of those in $C_r$, while the opposite holds for Agent~2. We call $C_{j^*}$ the \emph{disputed commodity} because it is the only one that in equilibrium may be possibly (but not necessarily) allocated to both agents.

To characterize the quantities $x_{j^*}$ and $y_{j^*}$ of the disputed commodity $C_{j^*}$ allocated to agents~1 and~2 in equilibrium, define the parameter 
\begin{equation}\label{eq1}
\xi = \frac{1}{2}\left[\frac{\nu_2\left(C_r\right)}{\nu_2\left(C_{j^*}\right)}-\frac{\nu_1\left(C_\ell\right)}{\nu_1\left(C_{j^*}\right)}+1\right].
\end{equation}
Arranged by cases, we obtain
\begin{equation}\label{eq2}
x_{j^*} = \begin{cases}
0 & \text{if } \xi \leq 0,\\
\xi\omega(C_{j^*}) & \text{if } 0 < \xi < 1, \\
\omega(C_{j^*}) & \text{if } \xi \geq 1,
\end{cases}
\end{equation}
with $y_{j^*}=\omega(C_{j^*})-x_{j^*}$.

A simple exercise looks at how the equilibrium utility $V_1^*(\pi)$ of agent $1$ changes when the classification $\pi$ is perturbed. Because $V_1^*(\pi)$ depends only on the allocation of the disputed commodity (up to null sets), the relevant information is captured by the derivatives of $V_1^*$ with respect to the two extreme points of $C_{j^*}= \left(\theta_1,\theta_2\right)$. For instance, a small change in $\theta_1$ entails a change in the equilibrium utility for Agent~1 given by
$$
\frac{\partial V_1^*(\pi)}{\partial \theta_1}=
\begin{cases}
f_1(\theta_1) & \text{if } \xi\leq 0,\\
\ds \frac{\nu_2(C_r)}{2\nu_2^2(C_{j^*})}\left[ \strut \nu_1(C_{j^*})f_2(\theta_1) -f_1(\theta_1)\nu_2(C_{j^*}) \right] & \text{if } 0<\xi<1,\\
0 & \text{if } \xi\geq 1,
\end{cases}$$
and this holds regardless of how $\pi \setminus\{C_{j^*}\}$ is defined, or perturbed.

For a different exercise, suppose $f_1(t)=2(1-t)$ and $f_2(t)=2t$ and consider the market design problem of finding the two-commodity classification that maximizes the sum of agents' utilities in equilibrium. By setting a threshold $\eta$ in $(0,1]$, the designer chooses the classification $\pi (\eta)$ formed by the two commodities $A=[0,\eta)$ and $B=[\eta,1]$.

Because the leveling point is at $\theta = \nicefrac{1}{2}$, the disputed commodity is $B$ for $\eta \le \nicefrac{1}{2}$, and $A$ otherwise. Using (\ref{eq1}) and (\ref{eq2}), we compute the utilities $V^*_1 (\eta)$ and $V^*_2 (\eta)$ that the two agents receive in equilibrium as functions of $\eta$:  
$$
V^*_1 (\eta) = \left\{ 
\begin{array}{l} 
\frac{1}{2} \\[1mm]
2\eta-\eta^2 \\[1mm]
\frac{2-\eta}{2\eta}
\end{array} 
\right.
\text{ and }\quad
V^*_2 (\eta) = \left\{ 
\begin{array}{l} 
\frac{1+\eta}{2(1-\eta)} \\[1mm]
1-\eta^2 \\[1mm]
\frac{1}{2}
\end{array} 
\right.
\quad\text{ for }
\left\{
\begin{array}{l} 
0\leq\eta\leq 1-\frac{\sqrt{2}}{2} \\[1mm]
1-\frac{\sqrt{2}}{2}< \eta <  \frac{\sqrt{2}}{2} \\[1mm]
 \frac{\sqrt{2}}{2}  \leq\eta\leq 1\\[1mm]
\end{array}
\right. 
$$
The sum $V^*_1 (\eta) + V^*_2 (\eta)$ of the equilibrium utilities  is maximized at $\eta^*=\frac{1}{2}$.

\section{Refining classifications}

Refining a classification expands trading opportunities: goods once conflated in the original classification are acknowledged as distinct commodities. Formally, we say that a classification $\rho$ \emph{refines} a classification $\pi$ (and write $\rho\succ\pi$) if the latter belongs to the algebra generated by $\rho$. When $\rho\succ\pi$, every $\pi$-bundle corresponds to a unique $\rho$-bundle and every feasible exchange in $\mathcal{E}(\pi)$ can also be realized within $\mathcal{E}(\rho)$, but the set of tradable commodities in $\rho$ is larger. Therefore, allocations that are Pareto-optimal under the initial classification may cease to be so.

This section compares configurations achieved by increasing the number of commodities and refining the classification underlying the economy.  Our modest goal is to show that refinements may sometimes yield unexpected outcomes. We hope that this may be a first step towards more general results. We assume that goods are assigned only via competitive equilibria consistent with the given classification, except for Section~\ref{sec44} where the configuration need not be competitive. 

\subsection{Refinements may switch trading positions}

The introduction of a new commodity may drastically change individual trading positions. An agent who is a buyer for commodity $A$ may switch positions and become a seller for $A$ when the underlying classification is refined. A market designer who knows only agents' demands for a given classification (but not their preferences) may not even guess the direction of their trades under a finer classification.

We compare an economy based on three commodities $A$, $B$, and $C$, against one that refines the classification by splitting $C$ into two distinct commodities. There is an agent who consumes only commodity $A$ in the first economy but consumes only $B$ in the second one. This occurs even though neither $A$ nor $B$ are directly affected by the refinement and despite their relative prices staying identical.

\begin{example}
Consider a society where $\omega$ is the Lebesgue measure and there are $4$ agents with different claims: agents $1$ and $2$ have claim $\frac{1}{3}$, while $3$ and $4$ have claim $\frac{1}{6}$. Preferences are linear and respectively based on  the evaluation measures:
$$\nu_1(F)=\frac{3}{2}\omega\left(F\cap\left[0,\frac{2}{3}\right)\right), \qquad \nu_2(F)=\frac{3}{2}\omega\left(F\setminus\left[\frac{1}{3},\frac{2}{3}\right)\right),$$
$$\nu_3(F)=\frac{5}{3}\omega\left(F\cap\left[\frac{1}{3},\frac{2}{3}\right)\right)+\frac{8}{3}\omega\left(F\cap\left[\frac{2}{3},\frac{5}{6}\right)\right),$$
$$\nu_4(F)=\frac{5}{3}\omega\left(F\cap\left[\frac{1}{3},\frac{2}{3}\right)\right)+\frac{8}{3}\omega\left(F\cap\left[\frac{5}{6},1\right]\right).$$
Let $\pi=\{A,B,C\}$ be the classification with $A=\left[0,\frac{1}{3}\right)$, $B=\left[\frac{1}{3},\frac{2}{3}\right)$ and $C=\left[\frac{2}{3},1\right]$. 

When all commodities carry the same price, agent $1$ is indifferent between $A$ and $B$, agent $2$ between $A$ and $C$, while $3$ and $4$ prefer $B$ over the other commodities. At the only equilibrium in $\mathcal{E}(\pi)$, prices are identical and the agents are allocated the bundles
$$x^1=\left(\frac{1}{3},0,0\right),\ x^2=\left(0,0,\frac{1}{3}\right),\ x^3=x^4=\left(0,\frac{1}{6},0\right).$$

Refine $\pi$ into the classification $\rho=\{A,B,C^1,C^2\}$ by splitting $C$ into two intervals $C^1=\left[\frac{2}{3},\frac{5}{6}\right)$ and $C^2=\left[\frac{5}{6},1\right]$. Agents $1$ and $2$ value $C$ as much as $C_1$ or $C_2$. Instead, the refinement allows $3$ and $4$ to differentiate between which parts of $C$ they like more: $3$ prefers $C_1$ while $4$ prefers $C_2$. At the only equilibrium in $\mathcal{E}(\rho)$, the four commodities have identical prices, and the agents are allocated the bundles:
$$x^1=\left(0,\frac{1}{3},0,0\right),\ x^2=\left(\frac{1}{3},0,0,0\right),\ x^3=\left(0,0,\frac{1}{6},0\right),\ \ x^4=\left(0,0,0,\frac{1}{6}\right).$$
In the economy $\mathcal{E}(\pi)$ agent $1$ trades away her endowments of $B$ and $C$ and buys $A$, while in $\mathcal{E}(\rho)$ she trades away $A, C_1, C_2$ and buys only $B$. Nevertheless, the commodities $A$ and $B$ and their relative prices are the same in both economies. Moreover, 1's switch in trading positions leaves unchanged the utility of his allocation under either classification.
\end{example}

\subsection{Refinements may not be welfare-improving}

In many situations, increasing the number of commodities allows every agent to achieve higher levels of utility. However, it is possible that introducing a new tradable commodity favors some agents and damages others. A refinement may fail to be Pareto-improving or even welfare-improving, where the \emph{(utilitarian) social welfare} associated with a classification $\pi$ is defined as the sum of the utilities that agents receive in any competitive equilibrium of $\mathcal{E}(\pi)$. 

The next example describes a society where no refinement of the initial classification yields strictly higher social welfare. The initial classification has two commodities $A$ and $B$: we show that splitting $A$ into two (or more) commodities leaves the social welfare unchanged, while splitting $B$ strictly reduces it. Hence, creating new commodities (in any number) is never Pareto- nor welfare-improving.

\begin{example}\label{Ex.UW}
Let $\omega$ be the Lebesgue measure on $\mathcal{I}$. Consider a society where there are $2n$ agents with identical claims, arranged in two groups of equal size. Preferences are linear and based on the evaluation measures:
\begin{align*}
\nu_1(F) & = \omega\left(F\cap\left[0, \frac{1}{2}\right)\right) + \frac{3}{2}\omega\left(F\cap\left[\frac{1}{2}, \frac{3}{4}\right)\right)+\frac{1}{2}\omega\left(F\cap\left[\frac{3}{4},1\right]\right), \mbox{ and}\\
\nu_2(F) & =2\omega\left(F\cap\left[\frac{1}{2},1 \right] \right).
\end{align*}
Consider the classification $\pi=\{A,B\}$, with $A=\left[0,\frac{1}{2}\right)$ and $B=\left[\frac{1}{2},1\right]$. At the competitive equilibrium for the economy $\mathcal{E}(\pi)$, all the commodities have identical prices, and each agent $i$ is allocated $\frac{1}{n}$ units of commodity $A$ or $B$ if she is in group $1$ or in group $2$, respectively.

We claim that no refinement of $\pi$ can improve the social welfare, and splitting $B$ actually reduces it. Let $\rho\succ\pi$ be a refinement of any size. If $\rho$ splits $A$ into any number of commodities $A_1,\dots,A_m$ (while leaving $B$ untouched), then agents remain indifferent between all the $A_i$'s. The equilibrium is essentially unaltered, and the social welfare does not change.

Suppose now that $\rho$ also splits $B$ into smaller intervals $B_1,\dots,B_k$.  After renaming, assume that the $B_j$'s are ordered so that $i<j$ implies $s<j$ for every $s$ in $B_i$ and $t$ in $B_j$. We show by contradiction that, in any equilibrium in $\mathcal{E}(\rho)$, agents in group $1$ consume a positive amount of some $B_j$'s, which were all assigned to group $2$ under $\pi$. Because group $1$ evaluates the $B_j$'s less than group $2$, the sum of agents' utilities in $\mathcal{E}(\rho)$ must be strictly lower than in $\mathcal{E}(\pi)$.

Since in equilibrium the market must clear, agents of group $1$ must be allocated all the $A_i$'s. Because they are indifferent, this requires that each $A_i$ has the same price $p$, or otherwise they would only demand the cheapest one(s). But then the price of $B_1$ must be strictly higher than $p$, or otherwise agents in group $1$ would prefer to demand $B_1$ over any of the $A_i$'s. If only agents in group $2$ demand all the $B_j$'s, their prices must all equal the price of $B_1$ (because agents in $2$ are indifferent over the $B_j$'s), and hence will be strictly higher than $p$. But if agents in group $2$ sell their whole endowment of $A$ at a price $p$ and buy the same amount of $B$ at a strictly higher price, this violates the budget constraints.
\end{example}

The intuition behind this example is that when $\pi=\{A,B\}$, agents in group $2$ do not compete against agents in group $1$ and let the latter choose their best option. If a finer classification $\rho$ splits $B$, the agents in group $2$ compete for some goods that were previously consumed by agents of group $1$, even if the latter have a higher evaluation for those goods. The increased competition induced by the refinement may strictly reduce social welfare. 

For the stronger case where every refinement of the starting classification yields strictly lower social welfare, see Example~\ref{Ex.Bidimensional} in Section~\ref{sec51} of the Online Appendix, where we port our model to a two-dimensional space of goods. Section~\ref{A.2} in the online Appendix has two more related examples.

\subsection{The optimal number of commodities}\label{sec44}

Limitations on the number of tradable commodities may be driven by economic considerations. Operating within economies with many commodities is costly both for the agents, who have to process more information, and for the market infrastructure, that must handle more elaborate transactions. A market designer may prefer a simpler environment when the social cost of increasing the number of commodities is higher than the social benefit from a richer set of trading opportunities.

The optimal number of commodities depends on many subtle design issues and agents' preferences. As a first step, we consider the simple case where the configuration maximizes the social welfare net of a classification-related cost that is proportional to the number of commodities acknowledged by the classification.

Formally, let $SW(k)$ be the maximum social welfare that can be achieved using $k$ commodities {across any configuration $\langle \pi,(x^i)\rangle$ with $\vert\pi\vert\leq k$. We} do not assume that $(x^i)$ is a competitive allocation in $\mathcal{E}(\pi)$. The social cost of operating with $k$ commodities is $ck$, with $c\in(0,1)$. The designer faces the problem:
\begin{equation}\label{Problema}
\max_{k\geq 1} \> SW(k)-ck.
\end{equation}
where the function $SW$ is bounded from above, because each agent can attain at most utility $1$ by consuming all the goods available. Therefore, a solution $k^*$ to this problem always exists. Moreover, the optimal number of commodities $k^*$ is never larger than $\overline{k} = \frac{n-(1-c)}{c}$, because otherwise the net social welfare would be higher by choosing $k=1$ and giving all goods to one agent.

Without restrictions on agents' preferences, it is not possible to give tighter bounds than $1 \le k^* \le \overline{k}$. Consider the lower bound: if all evaluation measures equal $\omega$, then the sum of agents' utilities is constant for any configuration, and therefore it is optimal to minimize the number of commodities choosing $k^*=1$, for any $c > 0$. The next example exhibits an economy where the upper bound $\overline{k}$ is attained, for $c$ sufficiently small.

\begin{example}
Let $c \le \frac{1}{m+1}$ for some $m$ in $\mathbb{N}$. Assume that $\omega$ is the Lebesgue measure and let $\pi=\{A_0,\dots,A_{m^2-1}\}$ be a partition of $\cal I$ in $m^2$ intervals of identical size $\frac{1}{m^2}$. There are $n=m$ agents, who have linear preferences. The evaluation measure of agent $i$ is
$$\nu_i(F)=\frac{m^2}{m+1}\left[\sum_{j=1}^m\omega\left(F\cap A_{jm-i}\right)+\omega\left(F\cap A_{im-i}\right)\right];$$
Each agent~$i$ has a positive value only for the $m$ intervals $A_{m-i},A_{2m-i}, \dots,A_{m^2-1}$, and this is the same for each interval, except for $A_{im-i}$ that she values twice as much. Agents' evaluations have disjoint support, and between two intervals valuable for agent $i$ there exists at least an interval that another agent $j \ne i$ values twice as much.

Suppose momentarily $c=0$; because agents' evaluations have disjoint support, the highest social welfare is obtained by assigning to each agent the commodity he values the most; hence, the coarsest optimal classification is $\pi$. We argue that this allocation (attained using $\pi$) remains optimal if $c \le \frac{1}{m+1}$. Assume that one reduces the number of commodities in order to save on operating costs. Any classification with fewer commodities than $\pi$ has at least one element $B$ with size larger than $\frac{1}{m^2}$ that is valuable for at least two agents. We can adroitly split $B$ into two new commodities and reassign those to the agents who value them the most, increasing the social welfare by at least $\frac{1}{m+1}$, which is greater than the cost of introducing an additional commodity. Because the optimal partition $\pi$ has $m^2$ commodities, replacing $n=m$ and $c = \frac{1}{m+1}$ into $\overline{k} = \frac{n-(1-c)}{c} = m^2$ shows that the upper bound is attained.
\end{example}

\section{Conclusions}\label{sec5}

This work provides a theoretical framework to study how the allocation of resources within a society depends on the conflation of goods into commodities. We rely on three founding blocks: (i) there are constraints on which classifications are allowed; (ii) agents' preferences over bundles of goods induce their preferences over commodities; and (iii) goods are allocated according to some mechanism restricted to the commodities listed in the classification.

For simplicity, we added two assumptions: (i) commodities are disjoint intervals of the unit segment $[0,1]$; and (ii) commodities are allocated via competitive equilibria. Our general framework, however, may be studied with different restrictions on the classifications, or different allocation mechanisms.

\appendix
\section*{Appendix: proofs}

\setcounter{thm}{0} 

\begin{prop}
$(\Pi_{(\le k)},d_\omega)$ is a compact space where two classifications have zero distance if and only if they coincide up to null sets. 
\end{prop}
\begin{proof}
Let $h$ be the pseudo-metric $h(F,G)=\omega(F\triangle G)$ defined on the measurable sets over $\mathcal{I}$. Then $d_\omega(\pi,\rho)$ is the Hausdorff distance between the algebras $\sigma(\pi)$ and $\sigma(\rho)$. Then $d_\omega(\pi,\rho)=0$ if and only if $\sigma(\pi)$ and $\sigma(\rho)$ have the same closure; see Lemma 3.72, Aliprantis and Border, 2006.  Equivalently, every $B$ in $\sigma(\pi)$ corresponds to some $B^\prime$ in $\sigma(\rho)$ up to null sets. Because $\sigma(\pi)$ and $\sigma(\rho)$ are finite, this holds if and only if the two classifications $\pi$ and $\rho$ (respectively viewed as the generators for the two algebras) coincide up to null sets.

Let $\mathcal{J}$ be the class of intervals from $\mathcal{I} = [0,1]$. If we identify two intervals having zero-distance, then the function $\omega$ maps isometrically $\mathcal{J}_0=\{F\in\mathcal{J}:\,0\in F\}$ into a closed and bounded subset of $\mathcal{I}$, so that $(\mathcal{J}_0,h)$ is itself compact.
Therefore, $\Pi_{(\le k)}$ is compact because it is the image of the compact product space $\mathcal{J}_0^{k-1}$ under the continuous function
$$\varphi(F_1,\dots,F_{k-1})=\left\{F_{i+1}\setminus F_i:\, \omega(F_{i+1})>\omega(F_i)\text{ and }F_k=\mathcal{I}\right\}.$$
\end{proof}

\begin{thm}
The competitive {equilibria} correspondence $\mathcal{W}$ is compact-valued and upper-hemicontinuous on the compact space $(\Pi_{(\le k)},d_\omega)$.
\end{thm}
\begin{proof} We prove first the continuity and compactness of the equilibrium correspondence $\widehat{\mathcal{W}}$ over the class $\widehat{\Pi}_{k}$ of partitions formed by exactly $k$ subsets. (These subsets need not be intervals and may have zero measure.) Then we show that the classifications in $\Pi_{(\le k)}$ can be continuously identified with (equivalence classes of) partitions in $\widehat{\Pi}_ k$ and use this fact to ``export'' the continuity and compactness of $\widehat{\mathcal{W}}$ over $\widehat{\Pi}_{k}$ to $\mathcal{W}$ over $\Pi_{(\le k)}$.

Endow $\widehat{\Pi}_{k}$ with the same pseudo-metric $d_\omega$ of $\Pi_{(\le k)}$.  Any classification $\hat{\pi}$ in $\widehat{\Pi}_{k}$ defines an exchange economy $\mathcal{E}(\hat{\pi})$, where null sets correspond to degenerate commodities with zero endowment. Such economies $\mathcal{E}(\hat{\pi})$ satisfy the assumptions of the Theorem in Hildenbrand and Mertens (1972),\footnote{\ Their assumption $\gamma$ posits strictly positive initial endowments for each agent, but their proof uses only the weaker property that each agent's endowment is not a cheapest point. This latter condition is met in our economies.} and thus the equilibrium correspondence $\widehat{\mathcal{W}}$ is compact-valued and upper-hemicontinuous over $\widehat{\Pi}_{k}$. 

Recall that $\pi$, $\pi^\prime$ coincide up to null sets if and only if $d_\omega(\pi,\pi^\prime)=0$, in which case they essentially define the same exchange economy. Then $\widehat{\mathcal{W}}(\pi)$ and $\widehat{\mathcal{W}}(\pi^\prime)$ are essentially the same set of allocations, where the only relevant commodities correspond to the non-null subsets in $\pi$ and $\pi^\prime$.

Consider the map $\gamma$ that associates each classification $\pi$ in $\Pi_{(\le k)}$ to the set $\gamma(\pi)$ of all partitions in $\widehat{\Pi}_{k}$ that coincide with $\pi$ up to null sets. Because every partition $\hat{\pi}$ in $\gamma(\pi)$ has zero distance from $\pi$, the correspondence $\gamma: \Pi_{(\le k)} \twoheadrightarrow \widehat{\Pi}_{k}$ is obviously continuous. Moreover, by the argument above, ${\mathcal{W}}(\pi)$ is essentially identical to the compact set $\widehat{\mathcal{W}}(\hat{\pi})$, for any $\hat{\pi}$ in $\gamma (\pi)$. We conclude that $\mathcal{W}$ can be identified with the composition of the upper-hemicontinuous maps $\widehat{\mathcal{W}}$ and $\gamma$, and thus it is itself upper-hemicontinuous and compact-valued.
\end{proof}

\begin{cor}
The competitive utilities correspondence $\Psi_i$ is compact-valued and upper-hemicontinuous on the compact space $(\Pi_{(\le k)},d_\omega)$.
\end{cor}
\begin{proof}
By construction, for every $\pi$ in $\Pi_{(\le k)}$ the set $\Psi_i(\pi)$ consists of all $V_i(\pi,x^i)$ with $x^i$ ranging over the bundles assigned to agent $i$ by the allocations in $\mathcal{W}(\pi)$. Every $\Psi_i (\pi)$ is compact because it is the continuous image of a compact set. And $\Psi_i$ is upper-hemicontinuous because it is the composition of a continuous function and an upper-hemicontinuous correspondence.
\end{proof}

\begin{prop}
Suppose that agents are not all identical and that $k>1$. Then {there is a competitive configuration $\langle\hat{\pi},(\hat{x}^i)\rangle$ with $\hat{\pi}$ in $\Pi_{(\le k)}$ that Pareto-dominates every competitive configuration based on a classification in an open set $O \subset \Pi_{(\le k)}$.
} 
\end{prop}
\begin{proof}
When $k>1$ and agents do not have identical preferences, there is a configuration $\langle\hat{\pi},(\hat{x}^i)\rangle$ that every agent prefers to the trivial configuration based on the classification $\pi_0=\{\mathcal{I}\}$. Define the set $O=\left\{\pi\in\Pi_{(\le k)}:\,\Psi_i(\pi)<V_i(\hat{\pi},\hat{x}^{i})\text{ for every }i\leq n\right\}$, that is not empty because it contains $\pi_0$, and it is open by the upper-hemicontinuity of each $\Psi_i$. By construction, every competitive configuration based on a classification in $O$ is Pareto-dominated by $\langle\hat{\pi},(\hat{x}^i)\rangle$.
\end{proof}

\begin{thm}
The set of competitive configurations based on classifications in $\Pi_{(\le k)}$ has a Pareto-optimal configuration.
\end{thm}
\begin{proof}
The set of competitive configurations based on classifications in $\Pi_{(\le k)}$ coincides with the graph of the correspondence $\mathcal{W}$, which is upper-hemicontinuous and compact-valued by Theorem~\ref{thm.0}.
Let $V$ be the continuous function that associates each competitive configuration $\langle \pi,(x^i)\rangle$ with the sum $\sum V_i(\pi,x^i)$. By Lemma 17.30 in Aliprantis and Border (2006), $m(\pi)=\max\left\{V(\pi,(x^i)):\,(x^i)\in \mathcal{W}(\pi)\right\}$ defines an upper-semicontinuous function on the compact set $\Pi_{(\le k)}$, and hence it has a maximum. Let $\langle \hat{\pi},(\hat{x}^i)\rangle$ be a competitive configuration that achieves such maximum. Then $\sum V_i(\hat{\pi},\hat{x}^i)\geq \sum V_i(\rho,y^i)$ for any competitive configuration $\langle \rho,(y^i)\rangle$ with $\rho$ in $\Pi_{(\le k)}$. We conclude that $\langle \hat{\pi},(\hat{x}^i)\rangle$ is Pareto optimal.
\end{proof}

\begin{prop}
If $\langle\pi,(x^i)\rangle$ is a Pareto-optimal configuration such that $x^i_C>0$ for every $i$ and every $C$ in $\pi$, then one can redefine agents' claims so that $\langle\pi,(x^i)\rangle$ is competitive.
\end{prop}
\begin{proof}
The second Welfare Theorem applied to the allocations in the economy $\mathcal{E}(\pi)$ implies that there exists a price system $p$ in $\mathbb{R}^\pi_+$ that supports the allocation $(x^i)$. Then $\langle \pi,(x^i)\rangle$ is a competitive configuration after we redefine agents' claims as:
$$\kappa_i =\sum_{B \in \pi}  p_B\frac{x^i_B}{\omega(B)}.$$
\end{proof}

\newpage
\setcounter{page}{1}
\renewcommand*{\thesubsection}{\arabic{subsection}}
\setcounter{equation}{0}

\begin{center}
{\large \bf Market allocations under conflation of goods}\\[4mm]

{\LARGE \tt Online Appendix}\\[2mm]

\url{https://mlicalzi.github.io/research/MACG-Online-Appendix.pdf}
\end{center}

\subsection{Pareto-dominated configurations}\label{A.1}

We show one way to formulate Proposition~\ref{prop:OpenSet} in measure-theoretic terms. Intuitively, this requires that the endowment of goods is well-distributed, in the sense that the measure $\omega$ is non-atomic. Let $\Pi_{(= k)}$ be the set of the classifications using \emph{exactly} $k$. Given $\pi$ in $\Pi_{(= k)}$, rearrange it in the string of intervals $(F_1,\dots,F_k)$ 
where $i<j$ if and only if $s<t$ for every $s$ in $F_i$ and $t$ in $F_j$. If we associate $\pi$ with the vector $\omega(\pi) = \left(\omega(F_1),\dots,\omega(F_k)\right)$ in $\mathbb{R}^k$, then the map $\pi \mapsto \omega(\pi)$ is continuous and injective. Define the measure $\lambda(O)$ for a set $O \subset \Pi_{(= k)}$ as the Lebesgue measure of  the set $\{\omega(\pi):\,\pi\in O\}$. 

\begin{cor}\label{cor:NonZeroSet}
Suppose that agents are not all identical and $k>1$. Then there is a non-null subset $O$ of classifications in $\Pi_{(= k)}$ whose competitive configurations are Pareto-dominated by some competitive configuration $\langle\hat{\pi},(\hat{x}^i)\rangle$ with $\hat{\pi}$ in $\Pi_{(= k)}$.
\end{cor}

The following example illustrates a society with two agents, where every configuration can be improved both from a Paretian and a utilitarian point of view with a suitable refinement of the underlying classification.

\begin{example}\label{Pathological}
Consider an economy where $\omega$ is the Lebesgue measure, and there are two agents with identical claims. Let $S\subset\mathcal{I}$ denote the Smith-Volterra-Cantor set (SVC set for short), which is a measurable set of size $\frac{1}{2}$ with the property that every non-null interval in $\mathcal{I}$ contains a non-null interval disjoint from $S$; see the $\epsilon$-Cantor set in Aliprantis and Burkinshaw (1981, p.~141). Agents of each group have linear preferences based on the evaluation measures:
$$\nu_1(F)=2\omega(F\cap S), \quad \nu_2(F)=2\omega(F\setminus S).$$
We claim that the only Pareto-optimal configurations assign the $0$ bundle to all agents of group $1$.

Let $\langle\pi,(x^i)\rangle$ be a configuration and let the interval $B\in\pi$ be a commodity such that $x_B^1>0$. By the properties of the SVC set $S$, there exists an interval $C\subseteq B$ such that $C\cap S=\emptyset$, and so $\nu_1(C)=0$ and $\nu_2(C)=2\omega(C)$. If we label $C$ as a new commodity, we obtain 
a finer classification $\rho$ under which one can transfer all goods of type $C$ previously assigned to $1$ to agent $2$, while leaving the rest of the allocation unchanged. But this benefits agent $2$ without causing harm to $1$ (because her evaluation of $C$ is null), proving that $\langle\pi,(x^i)\rangle$ is Pareto-dominated.

Formally, let $B^1$ and $B^2$ be the two (possibly empty) intervals obtained by removing $C$ from $B$. Let $\rho$ be a refinement of the classification $\pi$, where the commodity $B$ has been replaced with $B^1,B^2$ and $C$. Consider a new allocation $(y^i)$ in $\mathcal{E}(\rho)$ where the bundle assigned to agent~1 is
$$y^1_A=\frac{\omega(A)}{\omega(B)}x_B^1\text{ if }A\in\{B^1,B^2\},\ \ \ y^1_C=0,\ \ \ y^1_A=x^1_A\text{ otherwise;}$$
and the bundle assigned to agent $2$ is:
$$y^2_A=\frac{\omega(A)}{\omega(B)}x_B^2\text{ if }A\in\{B^1,B^2\},\ \ \  y^2_C=\frac{\omega(C)}{\omega(B)}x_B^2+\frac{\omega(C)}{\omega(B)}x_B^1,\ \ \  y^2_A=x^2_A\text{ otherwise}.$$
Computations shows that $(y^i)$ is a feasible allocation in $\mathcal{E}(\rho)$ that agent $1$ finds equivalent to $(x^i)$, while agent $2$ strictly prefers it to $(x^i)$. Standard arguments based on the continuity and monotonicity of the function $V_i(\rho,\cdot)$ prove that one can modify $(y^i)$ into a new allocation that every agent strictly prefers to $(x^i)$. 
\end{example}
Clearly, Example~\ref{Pathological} relies crucially on the assumption that agents' evaluations of goods are expressed through extremely elaborated subsets of $\mathcal{I}$ (such as the SVC set) while commodities can only be defined as intervals. If we allow commodities to be arbitrary subsets of $\mathcal{I}$, then the classification $\pi=\{S,S^c\}$ would generate a Pareto-optimal configuration where all goods of type $S$ are assigned to agent $1$ and the rest to agent $2$. This suggests that the stronger the exogenous constraints on the classification of goods into commodities, the further agents may be from reaching optimal allocations.

\subsection{Comparative statics with opposed preferences}\label{app-opp}

Given the classification $\pi=(C_1,\dots,C_k)$ where intervals are naturally ordered, let  $p=(p(C_i))_i$ denote the system of competitive equilibrium prices, with $x=(x(C_i))_i$ and $y=(y(C_i))_i$ being the equilibrium bundles respectively assigned to $1$ and $2$. We outline the main steps, including the explicit computation of $\frac{\partial V_1^*(\pi)}{\partial \theta_1}$. 

\paragraph{1.} 
The ratio $\nu_1(C_i)/\omega(C_i)$ decreases as $i$ increases. Indeed, let $F(x)=\int_0^x f(t)\,dt$, so that $F'(x)=f(x)$. Assume $a<b<c$; given $A=(a,b)$ and $B=[b,c)$, by the Mean Value theorem $\nu_1(A)/\omega(A)=(F(b)-F(a))/(b-a)=f(\theta_A)$ for some $\theta_A$ in $A$, and similarly $\nu_1(B)/\omega(B)=(F(c)-F(b))/(c-b)=f(\theta_B)$ for some $\theta_B$ in $B$. When $f$ is decreasing, $f(\theta_A)>f(\theta_B)$ and thus $\nu_1(A)/\omega(A)>\nu_1(B)/\omega(B)$. Specularly, $\nu_2(C_i)/\omega(C_i)$ increases as $i$ increases.

\paragraph{2.} 
Suppose that $x$ is an equilibrium allocation. If 
$$\frac{p(C_j)}{p(C_i)} > \dfrac{\> \nu_1(C_j)/ \omega(C_j) \>}{\nu_1(C_i)/ \omega(C_i)},$$
then $x(C_j)=0$. Correspondingly, $x(C_i)>0$ and $x(C_j)>0$ requires that the above expression holds as an equality. Similar considerations hold for $y$ and $\nu_2$.
 
\paragraph{3.} 
There is a $j^*$ such that $x(C_i)=\omega(C_i)$ for $i<j^*$ and $y(C_j)=\omega(C_j)$ for $j>j^*$.
\begin{proof}
Suppose $i<j$ and $x(C_j)>0$. Then (2) gives
$$\frac{p(C_j)}{p(C_i)} \le \dfrac{\> \nu_1(C_j)/ \omega(C_j) \>}{\nu_1(C_i)/ \omega(C_i)} < 1,$$
where the last inequality follows from (1). Using (1) again gives
$$\frac{p(C_j)}{p(C_i)} < 1 < \dfrac{\> \nu_2(C_j)/ \omega(C_j) \>}{\nu_2(C_i)/ \omega(C_i)}$$
and thus $y(C_i)=0$, or (2) would be violated. Therefore, because the market clears in equilibrium, $x(C_i)=\omega(C_i)$.
\end{proof}

\paragraph{4.} 
Let $C^*=C_{j^*}$ be the disputed commodity; define $C_\ell=\bigcup_{i<j^*}C_i$ and $C_r=\bigcup_{j>j^*}C_j$. If $x(C^*)>0$ and $y(C^*)>0$, then $x(C^*)=\xi\omega(C^*)$, where
\begin{equation}\label{eq-xi}
\xi=\frac{1}{2}\left[\frac{\nu_2(C_r)}{\nu_2(C^*)}-\frac{\nu_1(C_\ell)}{\nu_1(C^*)}+1\right]. \tag{*}
\end{equation}
\begin{proof}
By (1), we have $x(C_i)=\omega(C_i)$ for $i<j^*$ and $x(C_j)=0$ for $j>j^*$. Applying (2) gives
$$\frac{p(C^*)}{p(C_i)} = \dfrac{\> \nu_1(C^*)/ \omega(C^*) \>}{\nu_1(C_i)/ \omega(C_i)} \quad\text{ for every }i<j^*,$$
from which we obtain
$$\omega(C_i)p(C_i)=\left[ \frac{\omega(C^*)p(C^*)}{\nu_1(C^*)} \right] \nu_1(C_i) \quad\text{ for every }i<j^*.$$
Given the system of prices $p$, the worth of the bundle $x$ for Agent~1 is:
$$\sum_{i<j^*}\omega(C_i)p(C_i)+\xi \omega(C^*) p(C^*) = \left[ \frac{\omega(C^*)p(C^*)}{\nu_1(C^*)}\right] \cdot \left[\strut \nu_1(C_\ell)+\xi\nu_1(C^*)\right].$$
Similarly, the worth of the bundle $y$ for Agent~2 is:
$$\sum_{j>j^*}\omega(C_j)p(C_j)+(1-\xi)\omega(C^*) p(C^*) = \left[ \frac{\omega(C^*)p(C^*)}{\nu_2(C^*)}\right] \cdot \left[\strut \nu_2(C_r)+(1-\xi)\nu_2(C^*)\right],$$
Because in equilibrium $x$ and $y$ must have the same worth at $p$, we have:
$$\frac{\nu_1(C_\ell)+\xi\nu_1(C^*)}{\nu_1(C^*)}=\frac{\nu_2(C_r)+(1-\xi)\nu_2(C^*)}{\nu_2(C^*)}$$
from which (\ref{eq-xi}) follows.
\end{proof}

\paragraph{5.} 
By a standard continuity argument:
$$
x_{j^*} = \begin{cases}
0 & \mbox{if } \xi \leq 0,\\
\xi\omega(C_{j^*}) & \mbox{if } 0 < \xi < 1, \\
\omega(C_{j^*}) & \mbox{if }  \xi \geq 1,
\end{cases}
$$
with $y_{j^*}=\omega(C_{j^*})-x_{j^*}$.

\paragraph{6.} 
Suppose $C^*=(\theta_1,\theta_2)$ and denote by $V_1$ the utility that agent $1$ obtains from the bundle $x$. Using (5), we have
$$V_1^*=\begin{cases}
        \ds \nu_1(C_\ell) & \text{ if }\xi\leq 0,\\
        \ds \nu_1(C_\ell)+\xi\nu_1(C^*) & \text{ if }0<\xi<1,\\
        \ds \nu_1(C_\ell\cup C^*) &\text{ if }\xi\geq 1.
    \end{cases}$$
In particular, when $0 < \xi < 1$, substituting (\ref{eq-xi}) from (4) gives:
$$V_1^* = \frac{\nu_1(C_\ell)+\nu_1(C^*)}{2}+\frac{\nu_1(C^*)\nu_2(C_r)}{2\nu_2(C^*)}=\frac{\nu_1(C_\ell\cup C^*)}{2}+\frac{\nu_1(C^*)\nu_2(C_r)}{2\nu_2(C^*)}.$$
Recall that $C_\ell=[0,\theta_1]$, $C^*=(\theta_1,\theta_2)$ and $C_r=[\theta_2,1]$. Therefore:
$$\frac{\partial \nu_1(C_\ell\cup C^*)}{\partial \theta_1}=0,\qquad \frac{\partial \nu_1(C^*)}{\partial\theta_1}=-f_1(\theta_1),\qquad \frac{\partial \nu_2(C^*)}{\partial \theta_1}=-f_2(\theta_1).$$
Then the derivative of $V^*_1$ with respect to $\theta_1$ when $0<\xi<1$ is:
$$\frac{\partial V_1^*}{\partial \theta_1}=
\ds \frac{\nu_2(C_r)}{2\nu_2^2(C_{j^*})}\left[ \strut \nu_1(C_{j^*})f_2(\theta_1) -f_1(\theta_1)\nu_2(C_{j^*}) \right].
$$

\subsection{Refinements may not be welfare-improving}\label{A.2}

The next example {exhibits an economy and} a classification $\pi$ with the following property: for every (finer) classification $\rho$ that splits a commodity from $\pi$ into two commodities, there is an agent who strictly prefers every competitive allocation in $\mathcal{E}(\pi)$ to any competitive allocation in $\mathcal{E}(\rho)$. In short, adding a new commodity damages at least one agent and therefore is not a Pareto-improvement for the society.

\begin{example}\label{Pareto-refinement}
Consider an economy where $\omega$ coincides with the Lebesgue measure. There are $4$ agents with identical claims and linear preferences based on the evaluation measures:
$$\nu_1(F)=2\omega\left(F\setminus\left[\frac{1}{4},\frac{3}{4}\right]\right),\ \ \nu_2(F)=2\omega\left(F\cap \left[\frac{1}{4},\frac{3}{4}\right]\right),$$
$$\nu_3(F)=2\omega\left(\left[0,\frac{1}{2}\right]\right),\ \ \nu_4(F)=2\omega\left(\left[\frac{1}{2},1\right]\right).$$
Let $\pi$ be the classification formed by the two intervals $A=\left[0,\frac{1}{2}\right]$ and $B=\left(\frac{1}{2},1\right]$. In the exchange economy $\mathcal{E}(\pi)$, agent $3$ cares only about commodity $A$, agent $4$ only about $B$, and agents $1$ and $2$ are indifferent between them. An equilibrium is achieved when the two commodities have the same price and agents demand, for example, the $\pi$-bundles:
$$x^1=x^3=\left(\frac{1}{4},0\right),\ x^2=x^4=\left(0,\frac{1}{4}\right).$$
We claim that for every refinement $\rho$ of $\pi$ formed by $3$ tradable commodities, there is an agent that strictly prefers $(x^i)$ to any competitive allocation in $\mathcal{E}(\rho)$. Precisely, we assume that $\rho$ is obtained by splitting $A$ into two commodities $A_1$ and $A_2$ and we prove that, in equilibrium, agent $3$ cannot afford $\frac{1}{4}$ units of goods of type $A_1$ or $A_2$, implying that $3$ receives a strictly lower utility under $\rho$. The same strategy shows that if $\rho$ is obtained by splitting $B$, then agent $4$ strictly prefers $\pi$ to $\rho$.

Assume $t\in\left(0,\frac{1}{2}\right)$ such that $\omega(A_1)=t$ and $\omega(A_2)=\frac{1}{2}-t$.
Let $p$ be a competitive price in $\mathcal{E}(\rho)$ normalized so that $p(B)=1$ and let $w$ be agent $3$'s wealth at $p$. We assume that $p(A_1)\leq p(A_2)$ (the other case is treated identically) so that agent $3$ demands exactly:
$$\frac{w}{p(A_1)}=\frac{1}{4}\left[t+\frac{p(A_2)}{p(A_1)}\left(\frac{1}{2}-t\right)+\frac{1}{2p(A_1)}\right]$$
units of commodity $A_1$.

Let us assume by contradiction that $w/p(A_1)$ is greater than $\frac{1}{4}$. There are two possible cases:
\begin{itemize}
	\item if $p(A_1)=p(A_2)\leq 1$, then each of the agents $1$, $2$ and $3$ demands $\frac{1}{4}$ units of commodity $A_1$ or $A_2$. This creates an excess of demand, and thus $p$ cannot be an equilibrium price. On the other hand, if $p(A_1)=p(A_2)>1$ then $w/p(A_1)$ is strictly less than $\frac{1}{4}$.
	\item If $p(A_1)<p(A_2)$ then agents $1$ and $3$ demand $A_1$ instead of $A_2$. Therefore, $p(A_2)\leq 2$, or no agents would demand $A_2$. At the same time, it must be that $p(A_1)\geq\frac{1}{2t}$ or agent $1$ would demand only  $A_1$, leaving $3$ with strictly less than $\frac{1}{4}$ units of $A_1$. Combining these two inequalities we obtain:
$$
\frac{w}{p(A_1)}=\frac{1}{4}\left[t+\frac{p(A_2)}{p(A_1)}\left(\frac{1}{2}-t\right)+\frac{1}{2p(A_1)}\right]\leq\frac{1}{4}\left[t+2t(1-2t)+t\right]=t-t^2
$$
which is strictly smaller than $\frac{1}{4}$ for every $t<\frac{1}{2}$.
\end{itemize}
\end{example}
The above example is based on refinements of $\pi$ formed only by $3$ intervals. If we allow for richer classifications, then we can find refinements of $\pi$ that are strictly preferred to $\pi$ by every agent in the society. As a way of illustration, let $\rho$ be formed by the intervals:
$$A=\left[0,\frac{1}{4}-\varepsilon\right),\ \ B=\left[\frac{1}{4}-\varepsilon,\frac{1}{2}\right],\ \ C=\left(\frac{1}{2},\frac{3}{4}-\varepsilon\right],\ \ D=\left(\frac{3}{4}-\varepsilon,1\right]$$
with $\varepsilon\in\left(0,\frac{1}{4}\right)$. For $\varepsilon$ sufficiently small, an equilibrium in $\mathcal{E}(\rho)$ is achieved when all commodities have identical prices and each agent consumes the whole of a commodity ($1$ gets $A$, $2$ gets $C$, $3$ gets $B$, and $4$ gets $D$). This leaves every agent with a utility strictly larger than the one they received with the allocation $(x^i)$.

The following example refines both Example \ref{Ex.UW} in the main text and Example~\ref{Pareto-refinement} above by describing an economy where every refinement of the starting classification gives a strictly lower social welfare. The setup is similar to Example~\ref{Ex.UW}, but the set of feasible classifications is curtailed by assuming that commodity B is an atom, so that some tradable commodities cannot be split into smaller parts.

\begin{example}\label{Utilitarian-refinement}
Let $\lambda$ be the Lebesgue measure on $\mathcal{I}$ and $\delta_{\{1\}}$ denote the Dirac measure for the singleton $\{1\}$. We consider a society where there are $2n$ agents with identical claims, and the measure $\omega$ is given by:
$$\omega(F)=\frac{1}{2}\left(\lambda(F)+\delta_{\{1\}}(F)\right).$$
There are only two types of agents, forming groups of equal size. Agents have linear preferences based on the evaluation measures:
$$\nu_1(F)=\lambda(F) \quad and\quad \nu_2(F)=\frac{1}{4}\lambda\left(F\cap\left[0,\frac{1}{2}\right]\right)+\frac{3}{4}\lambda\left(F\cap \left[\frac{1}{2},1\right]\right)+\frac{1}{2}\delta_{\{1\}}(F).$$
Intuitively, agents of type $1$ value all types of goods identically, while those of type $2$ care more about goods in $\left[\frac{1}{2},1\right)$ and especially about those labeled with $1$.

Let $\pi$ be the classification formed by the commodities $A=\left[0,1\right)$ and $B=\{1\}$. At the competitive equilibrium, $A$ and $B$ have the same prices, with every agent from group $1$ consuming $\frac{1}{2n}$ units of commodity $A$ and every agent from group $2$ consuming $\frac{1}{2n}$ units of $B$. 

We prove that, if $\rho\succ\pi$, then every competitive allocation in $\mathcal{E}(\rho)$ assigns a positive amount of goods of type $A$ to agents in group $2$. Because the utility received from goods of type $A$ is higher for agents in group $1$, this implies that the sum of agents' utilities in $\mathcal{E}(\rho)$ must be strictly lower than in $\mathcal{E}(\pi)$. 

Suppose by contradiction that there exists a refinement $\rho$ of $\pi$ and a competitive allocation in $\mathcal{E}(\rho)$ such that agents in group $1$ consume all goods of type $A$ and those in group $2$ all goods of type $B$. Because $B$ is an atom, $\rho$ can refine $\pi$ only by splitting $A$ into smaller intervals and leaving $B$ intact. We write $\rho=\{A_1,\dots,A_m,B\}$ where $i<j$ implies $s<t$ for all $s\in A_i$ and $t\in A_j$. Because agents from group $1$ demand all commodities $A_1,\dots,A_m$, these must have all equal prices (otherwise, agents of group $1$ would demand only the cheapest ones). At the same time, $A_m$ must cost strictly more than $B$, or else agents in group $2$ would rather demand $A_m$ than $B$. Hence, the average price of the commodities $A_j$'s is strictly greater than the price of $B$, implying that each agent in group $2$ can demand more than $\frac{1}{2n}$ units of $B$. This leads to an excess of demand for $B$, which contradicts the assumption that prices are competitive.
\end{example}

\subsection{Bi-dimensional commodities}\label{sec51}

Consider an alternative description for the space of goods and its classifications. Suppose that each type of good is described by two values $x$ and $y$, where $x$ ranges over an interval $I$ and $y$ over an interval $J$. For instance, $x$ may indicate the position of the good and $y$ the date at which it is made available. The distribution of goods is a positive measure $\omega$ on $\mathcal{R}=I\times J$, normalized to $\omega(\mathcal{R})=1$. Bundles of goods are positive, $\omega$-integrable functions with the usual interpretation. Agents' preferences correspond to evaluation measures over $I\times J$.

An $\mathcal{R}$-classification is a partition $\pi=\left\{A\times B:\,A\in \pi_I,\,B\in\pi_J\right\}$, where $\pi_I$ and $\pi_J$ are finite partitions of $I$ and $J$ formed by intervals. Intuitively, $\pi$ may be seen as the result of cutting $\mathcal{R}$ by rows and columns drawn from $\pi_I$ and $\pi_J$. If we let $\Pi_{(\le k)}$ denote the set of $\mathcal{R}$-classifications with at most $k$ cells, this bi-dimensional setup includes our model as a special case, and all the results in Section~\ref{sec3} hold identically, with proofs easily mapped to the new setting.

However, this richer environment opens up more sophisticated considerations. As a way of illustration, consider a stronger version of Example \ref{Ex.UW}, where any component-wise refinement of $\pi = \pi_I \times \pi_J$ strictly decreases the social welfare. In short, adding either new rows or new columns is detrimental.

\begin{example}\label{Ex.Bidimensional}
Consider an economy where $\mathcal{R}=[0,1]\times[0,1]$ and $\omega$ is the Lebesgue measure on $\mathcal{R}$. Assume that $\mathcal{R}$ is partitioned in four equally sized quadrants:
$$A=\left[0,\frac{1}{2}\right) \times\left[0,\frac{1}{2}\right),\ B=\left[0,\frac{1}{2}\right)\times\left[\frac{1}{2},1\right],\ C=\left[\frac{1}{2},1\right]\times\left[\frac{1}{2},1\right],\ D=\left[\frac{1}{2},1\right]\times\left[0,\frac{1}{2}\right),$$
and further divide $B$ and $D$ in two more triangles by cutting along their diagonal:
 $$B^d=\left\{(x,y)\in B:\,x\leq 1-y\right\},\ B^u=\left\{(x,y)\in B:\,x>1-y\right\},$$
 $$D^d=\left\{(x,y)\in D:\,x\leq 1-y\right\},\ D^u=\left\{(x,y)\in D:\,x>1-y\right\}.$$
There are $2n$ agents with identical claims, arranged in two groups of equal size. Agents have linear preferences based on the evaluation measures
$$\nu_i(F)=\int_Ff_i(x,y)\,d(x,y),$$
computed using the densities
$$
f_1(x,y)=\begin{cases}
        1 & \text{ if }(x,y)\in A\cup C,\\
        \frac{3}{2} & \text{ if }(x,y)\in B^d\cup D^d,\\
        \frac{1}{2} & \text{ if }(x,y)\in B^u\cup D^u,
    \end{cases}
\quad\mbox{ and }\quad
f_2(x,y)=\begin{cases}
        2 & \text{ if }(x,y)\in B\cup D,\\
        0 & \text{ if }(x,y)\in A\cup C.
\end{cases}
$$
Consider the $\mathcal{R}$-classification $\pi=\{A,B,C,D\}$. The only competitive equilibrium in $\mathcal{E}(\pi)$ assigns commodities $A$ and $C$ only to agents in group $1$, and $B$ and $D$ only to those in group $2$. The sum of the equilibrium utilities is $\frac{3}{2}$.

Let $\rho$ be any refinement of $\pi$. Because $\rho$ obtains by splitting rows or columns in $\pi$, at least one of the quadrants $B$ and $C$ has to be partitioned into smaller rectangles. Suppose it is $B$ and let $B_1,\dots,B_m$ denote the rectangular commodities obtained from $B$. At least one $B_j$ is mostly contained below the diagonal, and $\omega(B_j\cap B^u)>\omega(B_j\cap B^u)$. Then agents in group $1$ evaluate $B_j$ more than $A$ or $C$, and yet their evaluation of $B_j$ is lower than that of agents in group $2$.

We claim that, for any equilibrium in $\mathcal{E}(\rho)$, agents in group $1$ entirely consume $A$ and $C$ and a positive fraction of the commodities obtained from either $B$ or $D$. Therefore, some goods initially assigned to group $2$ under $\pi$ are now allocated to $1$, with a reduction in total utility: for short, any refinement of $\pi$  induces a strict loss in the social welfare.

We only sketch the argument because it is a close analog of that in Example~\ref{Ex.UW}. In any equilibrium in $\mathcal{E}(\rho)$, agents in group~$1$ demand all the commodities generated by splitting $A$ or $C$, and so these must all have the same price $p$. At the same time, the price of commodity $B_j$ is strictly higher than $p$, or else agents in group~$1$ would demand $B_j$ over any commodity in $A \cup C$. Because agents in group~$2$ are indifferent among the commodities in $B \cup D$, they demand all of them only if they come at the same price as $B_j$, which is strictly higher than $p$. 

But this would imply that group $2$ can trade away its endowment of commodities of type $A$ and $C$ at the price $p$ and purchases the same amount of commodities of type $B$ and $D$ for a strictly higher price, in violation with the budget constraints.
\end{example}

This stronger version of Example~\ref{Ex.UW} exploits the property that every refinement of the $\mathcal{R}$-classification affects at least two commodities. The uni-dimensional model has no classification with such property.

\end{document}